\documentclass[conference]{IEEEtran}
\IEEEoverridecommandlockouts
\usepackage{cite}
\usepackage{amsmath,amssymb,amsfonts}
\usepackage{algorithmic}
\usepackage{graphicx}
\usepackage{textcomp}
\usepackage{xcolor}
\usepackage{booktabs,array,makecell,colortbl}
\usepackage{listings}
\usepackage{algorithm}
\usepackage{tcolorbox}
\usepackage{comment}
\usepackage{pifont}
\usepackage{url}
\usepackage[hidelinks]{hyperref}
\usepackage{tabularx}
\definecolor{lightgray}{gray}{0.95}
\definecolor{lightgreen}{rgb}{0.85,1.0,0.85}
\definecolor{lightred}{rgb}{1.0,0.85,0.85}
\usepackage{marvosym}

\def\BibTeX{{\rm B\kern-.05em{\sc i\kern-.025em b}\kern-.08em
    T\kern-.1667em\lower.7ex\hbox{E}\kern-.125emX}}

\begin{document}

\title{VulLink: A Dynamic Open-Access Vulnerability Graph Database for Cybersecurity Data Mining%
% \thanks{
% Project website:
% \href{http://34.129.186.158/}{VulLink Online Platform}.
% Source code:
% \href{https://github.com/L2-Izana/VulGD-Dynamic-Database}{VulLink GitHub Repository}.
% }
}
\author{
\IEEEauthorblockN{Luat Do}
\IEEEauthorblockA{\textit{Dept. of Computer Science and Information Technology} \\
\textit{La Trobe University}\\
Melbourne, Australia \\
21760427@students.latrobe.edu.au}
\and
\IEEEauthorblockN{Jiao Yin\textsuperscript{\Letter}\thanks{Corresponding author: Jiao Yin, jiao.yin@vu.edu.au.}}
\IEEEauthorblockA{\textit{Institute for Sustainable Industries and Liveable Cities} \\
\textit{Victoria University}\\
Melbourne, Australia \\
jiao.yin@vu.edu.au}

\and
\IEEEauthorblockN{Jinli Cao}
\IEEEauthorblockA{\textit{Dept. of Computer Science and Information Technology} \\
\textit{La Trobe University}\\
Melbourne, Australia \\
j.cao@latrobe.edu.au}
\and
\IEEEauthorblockN{Hua Wang}
\IEEEauthorblockA{\textit{Institute for Sustainable Industries and Liveable Cities} \\
\textit{Victoria University}\\
Melbourne, Australia \\
hua.wang@vu.edu.au}
}

\maketitle

\begin{abstract}
The rapid growth of software vulnerabilities has turned cyber threat intelligence analysis into a challenging data mining problem over heterogeneous and continuously changing sources. Public repositories such as the National Vulnerability Database (NVD), Common Vulnerabilities and Exposures (CVE), Common Weakness Enumeration (CWE), Exploit Database (EDB), and CVE Details provide valuable information, but their record-centric schemas make it difficult to capture cross-source relationships among vulnerabilities, weaknesses, exploits, affected products, vendors, and references. Existing graph-based vulnerability resources highlight the value of relational threat modelling, yet many remain static, offline, or difficult to access for downstream graph mining. This paper presents VulLink, a deployable, dynamic, open-source vulnerability graph database for cybersecurity data mining. VulLink integrates multiple public repositories through an automated Extract--Transform--Load (ETL) pipeline that converts isolated, record-centric vulnerability data into a continuously updated graph database with typed entities and explicit cross-source relationships. It provides an interactive Web interface and public API for exploring, querying, and exporting mining-ready vulnerability subgraphs. It also provides pre-computed embeddings of vulnerability descriptions generated by pretrained language models, which users can query and download by model and embedding dimension as semantic features for downstream mining tasks such as exploitability prediction. To demonstrate the practical utility of VulLink, we implement a downstream exploitability prediction use case that leverages heterogeneous graph context and semantic vulnerability features. The VulLink Source code and deployment materials are publicly available at \url{https://github.com/L2-Izana/VulGD-Dynamic-Database}.
% platform, including the Web interface, public API, source code, and deployment resources, is publicly available online\footnote{Project Web interface: \href{http://34.129.186.158/}{VulLink demonstration system}.
\end{abstract}


%------------------------------------------------------
\section{Introduction}
The increasing prevalence of software vulnerabilities poses significant threats to modern information systems, leading to data breaches, financial losses, and infrastructure disruptions \cite{yin2022cybersecurity,you2023knowledge}. Recent incidents illustrate the severity of this problem. The MOVEit breach in 2023 compromised data from more than 93 million individuals across critical sectors \cite{moveit2023}, while cyberattacks on Australian superannuation providers in 2025 caused extensive data theft and financial damage \cite{ausSuper2025}. At the same time, the volume of disclosed vulnerabilities continues to rise, with over 48,000 CVEs published in 2025 alone, an increase of 20\% over the previous year\footnote{\label{fn}\url{https://www.cvedetails.com}}.

To support vulnerability management, public repositories such as the National Vulnerability Database (NVD)\footnote{\url{https://nvd.nist.gov}}, Common Vulnerabilities and Exposures (CVE)\footnote{\url{https://cve.mitre.org}}, Common Weakness Enumeration (CWE)\footnote{\url{https://cwe.mitre.org}}, Exploit Database (EDB)\footnote{\url{https://www.exploit-db.com}}, and CVE Details\footnotemark[1] 
provide complementary information, including standardized identifiers, severity scores, weakness categories, exploit records, affected-product information, and external references \cite{yin2020apply}. However, these repositories are typically maintained as independent, record-centric sources rather than as a unified relational layer across vulnerability entities. As a result, important cross-source connections among vulnerabilities, weaknesses, exploits, products, vendors, and references are not always explicit or directly queryable, limiting the analysis of shared weaknesses, co-exploitation pathways, and software dependency chains for vulnerability risk assessment~\cite{sun2023CVEChainReasoning}. This fragmentation is also observed in practice, where 37\% of organizations report difficulties in remediation due to fragmented vulnerability intelligence \cite{swimlane2025}. Academic studies similarly report traceability and consistency gaps across cyber threat intelligence artifacts \cite{geras2024bigbeast}.

From a data mining perspective, this fragmentation creates a practical barrier. Before conducting graph mining, exploitability analysis, vulnerability clustering, or representation learning, researchers often need to collect data from multiple repositories, align identifiers, remove duplicated entities, extract relations, and construct task-specific graph datasets. This process is time-consuming and difficult to reproduce. The problem is further complicated by the dynamic nature of vulnerability intelligence: new CVEs, exploit records, affected-product mappings, and reference links are continuously released \cite{tang2020modelling}. Static datasets and one-off graph snapshots are useful for retrospective analysis, but they provide limited support for continuously updated cybersecurity analytics and benchmark construction.

Graph databases provide a natural way to represent vulnerability intelligence as connected entities and relations. Existing vulnerability knowledge graphs and cybersecurity graph systems have shown the value of modelling relationships among vulnerabilities, exploits, affected products, weaknesses, and external references \cite{kiesling2019sepseskg,yin2024vulkg,høst2023kg,yin2023empowering}. However, many existing resources are static, offline, task-specific, or not easily accessible through public interfaces. As a result, there remains a gap between fragmented public vulnerability repositories and the needs of researchers and practitioners who require dynamic, queryable, and reusable graph data for cybersecurity data mining.

To address this gap, we present VulLink, a dynamic and open-access vulnerability graph database for cybersecurity data mining. VulLink converts fragmented vulnerability records from multiple public repositories into a continuously updated graph database and exposes the resulting data through a Web interface and a public API. Specifically, this paper makes the following applied contributions:

\begin{itemize}
    \item \textbf{Dynamic multi-source vulnerability graph database construction.} We design and implement an automated Extract--Transform--Load (ETL) pipeline that integrates multiple public vulnerability repositories and converts isolated, record-centric data into a continuously updated graph database with typed entities and explicit cross-source relationships.
    \item \textbf{Open-access graph query and customizable subgraph export.} VulLink is a deployable Neo4j-backed graph database system that supports interactive graph exploration, click-based download of selected node and relationship data, and API-based customized subgraph retrieval through Cypher queries. This allows users to obtain task-specific vulnerability subgraphs for downstream data mining, including exploitability analysis, vulnerability clustering, multi-hop relation analysis, graph representation learning, and graph neural network benchmarking.
    \item \textbf{Pre-computed semantic features for downstream mining.} VulLink provides embeddings of vulnerability descriptions generated by pretrained language models. Users can select embeddings by model and embedding dimension, and download them together with the corresponding vulnerability attributes and graph context for downstream experiments. We further evaluate their utility through an exploitability prediction case study.
\end{itemize}
Together, these components make VulLink a practical and accessible graph data resource for cybersecurity data mining, supporting both exploratory analysis and downstream machine learning tasks.

% The remainder of the paper is structured as follows. Section~\ref{sec:relatedwork} discusses related work on vulnerability graphs, data pipelines, and LLM applications in cybersecurity. Section~\ref{sec:methodology} details the methodology and architecture of VulLink, describing each component of the system. Section~\ref{sec:demo_usecase} outlines the implementation details and configurations, and demonstrates the capabilities of VulLink through a practical use case. Finally, Section~\ref{sec:conclusion} concludes with a discussion on limitations and future research opportunities.

\section{Related Work}\label{sec:relatedwork}

Graph-based representations have been widely explored for cybersecurity analysis because they can explicitly model relationships among different entities, such as vulnerabilities, products, exploits, weaknesses, assets, and references \cite{xiangjie2018lotad,you2025learning,ngo2025optimizing}. Early systems such as CyGraph integrate heterogeneous cybersecurity information for situational awareness and graph-based reasoning \cite{noel2016cygraph}. Subsequent studies have constructed vulnerability knowledge graphs from public sources such as CVE, NVD, CWE, and Common Platform Enumeration (CPE)\footnote{\url{https://cpe.mitre.org/}} for vulnerability assessment, link prediction, risk analysis, and security reasoning \cite{wang2022software,yuan2021predicting}. Representative examples include SEPSES, which aggregates multiple cybersecurity feeds into an evolving knowledge graph \cite{kiesling2019sepseskg}; NVDText-KG, which extracts structured knowledge from NVD descriptions and applies graph embeddings for completion \cite{høst2023kg}; and VulKG, which models relationships among vulnerabilities, exploits, affected products, and domains for vulnerability risk analysis \cite{yin2024vulkg,yin2023vcbd}. These studies demonstrate the value of graph-structured vulnerability intelligence, but many existing resources remain static, task-specific, offline, or difficult to access through public interfaces.

Another relevant direction is dynamic cybersecurity data integration. Vulnerability intelligence changes continuously as new CVEs, exploit records, affected-product mappings, and external references are released. Public repositories such as NVD provide APIs and JSON feeds for accessing updated vulnerability records \cite{NVDDataJsonFeedsUpdateRule}, but these feeds are still organized around individual records rather than mining-ready graph structures. Recent studies have begun to treat knowledge graphs as evolving data systems, such as PageLLM for incremental security knowledge graph updating \cite{mistra2025pagellm} and Stream2Graph for online learning over large-scale dynamic knowledge graphs \cite{barry2022steam2graph}. 
However, these works do not directly provide an open, continuously updated vulnerability graph database with query and export functions for downstream cybersecurity data mining.

Textual vulnerability descriptions also provide important semantic information about attack conditions, affected components, impacts, and exploit patterns. 
Language models and text embeddings have been used in cybersecurity tasks such as vulnerability detection, threat intelligence extraction, and vulnerability exploitability prediction \cite{huang2024secbert,fieblinger2024actionablecyberthreatintelligence}. In vulnerability graph research, semantic representations have also been used for entity extraction, relation extraction, graph completion, and hidden-link discovery \cite{høst2023kg,xiang2025uncovering}. 
Recent studies have further investigated semantic and dynamic graph learning, including LLM-based knowledge graph link prediction \cite{li2024zero} and continual learning over evolving knowledge graphs \cite{ozbulak2025cast}. VulLink complements these method-oriented studies: it does not utilize language models to generate authoritative graph edges; instead, it provides a continuously updated vulnerability graph database with pre-computed description embeddings and customizable subgraph export for downstream cybersecurity data mining.

Table~\ref{tab:comparison} compares VulLink with representative vulnerability-focused graph systems. Here, semantic features refer to text-derived representations, such as embeddings or extracted semantic representations, that can support downstream mining. Compared with prior work, VulLink combines four practical properties that are important for applied cybersecurity data mining: dynamic updating, semantic features, multi-source coverage, and open Web/API access.

\begin{table}[htbp]
\centering
\caption{High-level comparison of VulLink with related systems.}
\label{tab:comparison}
\begin{tabular}{@{}l c c c c@{}}
\toprule
% \rowcolor{lightgray}
\textbf{System} & \makecell{\textbf{Dynamic} \\ \textbf{Pipeline}} & \makecell{\textbf{Semantic} \\ \textbf{Features}} & \makecell{\textbf{Source} \\ \textbf{Diversity}} & \makecell{\textbf{Open Web/} \\ \textbf{API Access}} \\
\midrule
\cite{yin2024vulkg} & \ding{55} & \ding{55} & \ding{51} & \ding{55} \\
\cite{kiesling2019sepseskg} & \ding{51} & \ding{55} & \ding{51} & \ding{55} \\
\cite{høst2023kg} & \ding{55} & \ding{51} & \ding{55} & \ding{55} \\
\cite{dulin2022multisourcegraph} & \ding{55} & \ding{55} & \ding{51} & \ding{55} \\
\cite{falcarin2024cybergraph} & \ding{51} & \ding{55} & \ding{51} & \ding{55} \\
VulLink & \ding{51} & \ding{51} & \ding{51} & \ding{51} \\
\bottomrule
\end{tabular}
\end{table}

\section{System Design}\label{sec:methodology}
\begin{figure*}[!t]
  \centering
  \includegraphics[width=2\columnwidth]{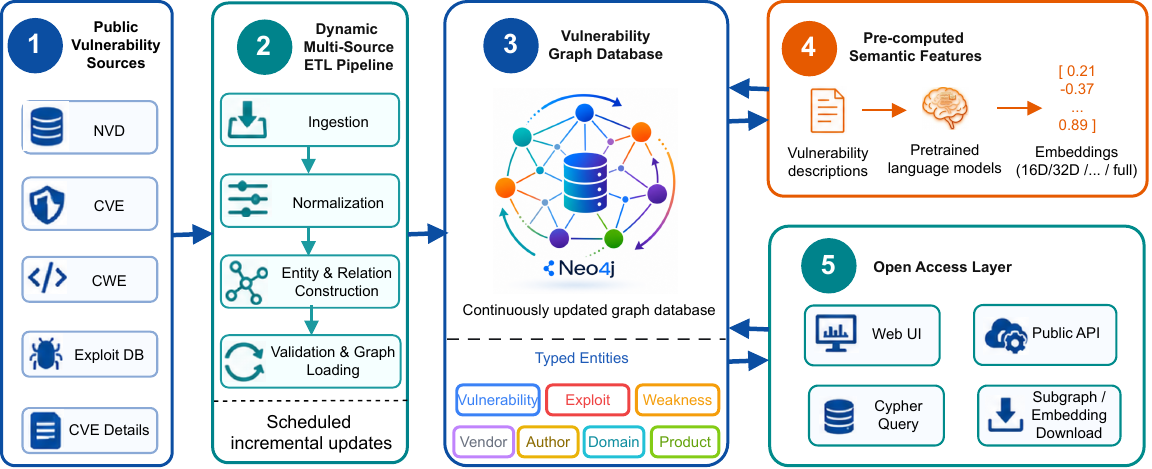} %[width=0.8\columnwidth]
  \caption{Overview of the VulLink system design. VulLink integrates public vulnerability repositories through a dynamic multi-source ETL pipeline, converts isolated records into a continuously updated graph database, provides pre-computed vulnerability-description embeddings, and exposes graph queries and subgraph/embedding downloads through both the Web interface and public API. 
  }
  \label{fig:VulLink_architecture}
\end{figure*}

This section presents the design of VulLink. As shown in Fig.~\ref{fig:VulLink_architecture}, VulLink is organized into five connected modules: public vulnerability sources, a dynamic multi-source ETL pipeline, a vulnerability graph database, pre-computed semantic features, and an open access layer. The overall design follows the main objective of VulLink: converting isolated, record-centric vulnerability data into a continuously updated graph database with typed entities, explicit cross-source relationships, downloadable semantic features, and public access for downstream cybersecurity data mining.

\subsection{Public Vulnerability Sources}

VulLink integrates multiple public vulnerability repositories that provide complementary information about software vulnerabilities. The current system uses NVD as the core vulnerability source and further incorporates CVE, CWE, ExploitDB, and CVE Details as supplementary sources. These sources differ in data format, update mechanism, entity granularity, and relationship coverage. For example, NVD provides structured CVE records, vulnerability descriptions, severity-related fields, configurations, references, and weakness mappings; CVE provides standardized vulnerability identifiers and metadata; CWE defines weakness categories; ExploitDB provides proof-of-concept exploit records and exploit authorship information; and CVE Details provides additional product, vendor, and vulnerability attribute information.

A key challenge is that these sources are not naturally organized as a unified graph. They expose records through source-specific schemas, APIs, feeds, or web-accessible data formats. The same vulnerability may appear across several repositories, while related entities such as exploits, products, weaknesses, vendors, and external references are often represented as fields, identifiers, URLs, or semi-structured text rather than as explicit graph objects. VulLink therefore treats each source as a separate input stream and uses the ETL pipeline to map source-specific records into a common graph representation.

\subsection{Dynamic Multi-Source ETL Pipeline}

The dynamic multi-source ETL pipeline is the core engineering component of VulLink. It is designed to support continuous data ingestion, source-specific preprocessing, entity and relation construction, validation, and graph loading. The main engineering challenge is that each public source exposes different identifiers, update formats, schema structures, and levels of relational completeness. VulLink therefore implements source-specific sub-pipelines while enforcing a shared graph schema at the loading stage. Instead of creating a one-time graph snapshot, VulLink updates the graph database through scheduled incremental updates, allowing newly available records to be merged into the existing graph without rebuilding the full database from scratch. As illustrated in Module 2 of Fig.~\ref{fig:VulLink_architecture}, the ETL pipeline contains four main stages.

\textbf{Data ingestion.} The ingestion stage retrieves vulnerability records from multiple public sources. Each source is handled by a source-specific sub-pipeline because the data formats and access patterns differ across repositories. For NVD, the pipeline processes structured vulnerability feeds and extracts CVE identifiers, descriptions, severity-related fields, weakness mappings, affected configurations, and references. For CWE, the pipeline retrieves weakness entries and maintains their mapping to vulnerabilities when such links are available. For ExploitDB, the pipeline extracts exploit records, exploit identifiers, exploit types, affected platforms, authors, and linked CVEs. For CVE Details, the pipeline collects additional product, vendor, and vulnerability attribute information. This modular design allows each source to be updated or extended independently.

\textbf{Preprocessing and normalization.} The preprocessing stage cleans and standardizes source records before graph construction. This includes parsing source-specific fields, 
% removing invalid or incomplete entries, 
standardizing identifiers such as CVE IDs and CWE IDs, normalizing product and vendor names when possible, and extracting domains from external reference URLs. The purpose of normalization here is not to force all sources into a single relational table, but to prepare heterogeneous records for conversion into typed graph entities and relationships.

\textbf{Entity and relation construction.} The transformed records are then mapped to graph nodes and edges. VulLink constructs typed entities such as \textit{Vulnerability}, \textit{Exploit}, \textit{Weakness}, \textit{Product}, \textit{Vendor}, \textit{Author}, and \textit{Domain}. It also constructs explicit relationships such as \texttt{EXPLOITS}, \texttt{AFFECTS}, \texttt{BELONGS\_TO}, \texttt{EXAMPLE\_OF}, \texttt{WRITES}, and \texttt{REFERS\_TO}. This step is central to VulLink because it converts source-specific fields and references into queryable cross-source relationships. For example, a CVE record, its affected products, related weakness categories, known ExploitDB entries, exploit authors, and external reference domains can be represented as a connected vulnerability subgraph.

\textbf{Validation and graph loading.} Before data are loaded into the graph database, VulLink checks whether the required identifiers and relationship endpoints are available. This validation step reduces the creation of orphaned nodes and malformed relationships. The graph loading stage uses idempotent Cypher operations, such as \texttt{MERGE}, to update the graph. This design allows the pipeline to repeatedly process newly downloaded or refreshed records while avoiding duplicated nodes and relationships. It also supports incremental updates: existing nodes can be matched and updated, while new entities and edges are inserted only when they are not already present.

The deployed pipeline is triggered by scheduled jobs. In the current implementation, the scheduling layer uses \texttt{Crontab} to run the update workflow automatically. Scheduled execution determines when the ETL process starts, while incremental update logic determines how new or modified source records are merged into the existing graph. This separation is important for maintaining a continuously updated graph database under limited server resources.

\subsection{Vulnerability Graph Database}

The output of the ETL pipeline is stored as a property graph. VulLink currently uses Neo4j as the graph database backend because it provides mature support for property-graph storage, graph traversal, Cypher queries, and integration with interactive visualization. Neo4j is an implementation choice; the main contribution of VulLink is the construction and deployment of a continuously updated vulnerability graph database rather than the use of a specific backend alone.

The graph schema follows the entity and relationship design of the VulKG framework \cite{yin2024vulkg}. As shown in Module 3 of Fig.~\ref{fig:VulLink_architecture}, VulLink represents major vulnerability intelligence entities as typed nodes. A \textit{Vulnerability} node is the central entity and is connected to other types of nodes through explicit relationships. An \textit{Exploit} node is connected to a vulnerability through \texttt{EXPLOITS}; a \textit{Product} node is connected through \texttt{AFFECTS}; a \textit{Weakness} node is connected through \texttt{EXAMPLE\_OF}; a \textit{Product} may be linked to a \textit{Vendor} through \texttt{BELONGS\_TO}; an \textit{Author} can be connected to an exploit through \texttt{WRITES}; and a \textit{Domain} node records external reference domains through \texttt{REFERS\_TO}.

This graph representation enables queries that are difficult to perform directly over isolated source records. Users can retrieve a vulnerability together with its affected products, related weaknesses, exploits, exploit authors, vendors, and reference domains in a single multi-hop query. The typed entity design also supports downstream graph mining tasks, including subgraph extraction, graph representation learning, vulnerability clustering, and exploitability prediction.

\subsection{Pre-computed Semantic Features} \label{sec:llm_embedding_strategy}

In addition to structured graph entities and relationships, VulLink provides semantic features derived from vulnerability descriptions. Vulnerability descriptions often contain information about attack conditions, affected components, impacts, and exploit patterns that may not be fully captured by structured fields. To make this information directly usable for data mining, VulLink generates pre-computed embeddings of vulnerability descriptions using pretrained language models.

The current implementation supports multiple embedding models, including \texttt{all-mpnet-base-v2}, \texttt{SecBERT}, and \texttt{facebook/fasttext-en-vectors}. The all-mpnet-base-v2 model is a sentence embedding model from the Sentence-Transformers framework \cite{song2020mpnet}. 
SecBERT is a cybersecurity-oriented pretrained BERT model released for security-text representation tasks \cite{secbert_repository}.
The FastText model provides a lightweight subword-based representation \cite{bojanowski2017enriching}. These models provide users with different trade-offs between domain specificity, representation capacity, and computational cost.

VulLink uses these models as semantic feature extractors rather than as authoritative knowledge generators. In other words, pretrained language models are not used to create factual graph edges. The graph structure is constructed from public vulnerability sources through the ETL pipeline, while embeddings are provided as additional downloadable features associated with vulnerability descriptions. This design reduces the risk of introducing language-model hallucinations into the graph database and keeps the factual graph layer separated from the learned semantic feature layer.

For each vulnerability with a valid description, VulLink generates embedding vectors and stores them in a server-side embedding store. Users can retrieve embeddings by specifying the model and desired dimension through either the Web interface or the public API. The system supports multiple embedding dimensions, including low-dimensional, moderate-dimensional, and full-dimensional versions, from 16D to 768D. The low- and moderate-dimensional versions are used for lightweight retrieval and downstream experiments, while full-dimensional embeddings are retained for higher-fidelity analysis. The detailed storage and dimensionality trade-offs are evaluated in Section~\ref{sec:LLMdemission}.

During regular updates, embeddings are generated or refreshed only for newly added vulnerability descriptions, while existing embeddings are reused from the embedding store. This avoids repeated inference over the full graph and keeps the semantic feature layer aligned with the incremental ETL process. When users request a specific embedding dimension, VulLink returns the pre-computed representation if available, or serves the closest reduced representation according to the configured retrieval policy. This pre-computation strategy avoids real-time language-model inference during user queries and reduces the burden on users who want to use vulnerability-description representations without rebuilding the embedding pipeline locally. In downstream experiments, these embeddings can be downloaded together with vulnerability attributes and graph context, enabling direct use in machine learning and graph mining workflows.

\subsection{Open Access Layer}

VulLink exposes the graph database and semantic features through an open-access layer. This layer provides two access channels, an interactive Web interface and a public API, both supporting Cypher-based graph queries as well as subgraph and embedding downloads. The purpose of this layer is to make the graph usable by both interactive users and automated data mining workflows.

The Web interface supports visual graph exploration. Users can inspect typed nodes and relationships, browse vulnerability subgraphs, view node properties, and explore common relationships without writing code. For advanced users, VulLink provides Cypher-based querying, allowing customized retrieval of task-specific subgraphs. For example, a user can query a vulnerability together with its affected products, vendor information, weakness category, exploit records, exploit authors, and external reference domains.

The public API supports programmatic access to the graph and embedding data. It allows users to download selected node types, export relationships with source and target information, submit customized Cypher queries, and retrieve pre-computed embeddings. This API-based design is important for reproducibility because researchers can define the exact graph extraction query used in an experiment and download the corresponding data directly. It also allows external systems to integrate VulLink into automated vulnerability analysis workflows.

Together, the Web interface and API allow VulLink to support different levels of use. Non-expert users can explore the graph visually, while researchers and practitioners can retrieve customized subgraphs and semantic features for downstream data mining. This open-access design is central to VulLink's role as a practical graph data resource rather than a closed or one-off vulnerability graph snapshot.

\section{Deployment and Evaluation}\label{sec:demo_usecase}
This section reports the deployed VulLink system and evaluates its practical utility through deployment configuration, graph scale, temporal coverage, embedding retrieval cost, and Web/API-based data access. We further include an exploitability prediction case study to examine whether VulLink's graph context and pre-computed semantic features can support downstream cybersecurity data mining.

\subsection{Deployment Configuration and Implementation Details}

\paragraph{Server environment.}
VulLink is deployed on a Virtual Private Server (VPS) with 2 vCPUs, 4GB RAM, 80GB SSD storage, and Ubuntu 24.04 LTS (x64). This setting reflects a cost-conscious deployment environment rather than a high-end research cluster, which is important for assessing whether VulLink can operate as a practical open-access data resource. The main resource pressure comes from graph loading, large query responses, and embedding retrieval, especially when users request high-dimensional semantic features.

\paragraph{Software stack.}
The system is containerized using Docker to improve deployment stability and environmental reproducibility. The graph database is implemented using Neo4j Community Edition 4.4.11. The dynamic ETL pipeline and API services are implemented in Python 3.10+. The pipeline uses the \texttt{transformers} package for language-model-based embedding generation, \texttt{scikit-learn} for dimensionality reduction, and FastAPI for programmatic access. Nginx is used as a reverse proxy for the Web service, while the React-based front end uses D3.js for interactive graph visualization. The dynamic ETL pipeline is triggered by scheduled jobs every two hours, following the update frequency of the NVD JSON feeds \cite{NVDDataFeeds}. To protect the deployed service from excessive real-time load, the interface and API enforce access limits for large embedding queries.

\paragraph{Back-end and front-end interaction.}
The back end coordinates the graph database, embedding store, ETL pipeline, and API gateway. Neo4j connections are established through the Bolt protocol, while the FastAPI layer handles tasks such as node export, relationship export, Cypher-based subgraph retrieval, and embedding download. The front end communicates with the graph database and the API layer to support both visual exploration and structured data retrieval. This separation allows VulLink to support interactive browsing, customized graph queries, and reproducible programmatic downloads without requiring users to deploy Neo4j or rebuild the ETL pipeline locally.

\subsection{Graph Snapshot Statistics and Temporal Coverage}

To characterize the scale and structural richness of VulLink, we report the latest snapshot of the graph. As shown in Table~\ref{tab:VulLink_stats}, the deployed graph contains 545,420 nodes and 1,660,599 relationships, including 343,598 vulnerability nodes, 46,605 exploit nodes, 962 weakness nodes, 97,684 product nodes, 27,575 vendor nodes, 10,155 author nodes, and 18,841 domain nodes. These entities are connected by multiple relationship types, including 675,377 \texttt{AFFECTS} edges, 71,054 \texttt{EXAMPLE\_OF} edges, 29,115 \texttt{EXPLOITS} edges, 87,866 \texttt{BELONGS\_TO} edges, 46,605 \texttt{WRITES} edges, and 750,582 \texttt{REFERS\_TO} edges.

\begin{table}[htbp]
\centering
\caption{Statistics of VulLink.}
\label{tab:VulLink_stats}
\begin{tabular}{@{}llll@{}}
\toprule
\textbf{Entity Label} & \textbf{Count} & \textbf{Relationship Type} & \textbf{Count} \\ \midrule
Vulnerability         & 343,598        & EXPLOITS                   & 29,115         \\
Exploit               & 46,605         & AFFECTS                    & 675,377        \\
Weakness              & 962            & BELONGS\_TO                & 87,866         \\
Product               & 97,684         & EXAMPLE\_OF                & 71,054         \\
Vendor                & 27,575         & WRITES                     & 46,605         \\
Author                & 10,155         & REFERS\_TO                 & 750,582        \\
Domain                & 18,841         &   -                         &  - \\
Total nodes  &  545,420 & Total edges & 1,660,599 
\\ \bottomrule
\end{tabular}
\end{table}

These statistics show that VulLink is not a small demonstration graph, but a deployable graph database at real-world vulnerability scale. The large number of cross-entity relationships is particularly important for data mining: vulnerabilities can be connected to affected products, vendors, weakness categories, exploit records, exploit authors, and external reference domains. This structure supports multi-hop relation analysis and subgraph extraction that would be difficult to conduct directly over isolated, record-centric repositories.

\begin{figure}[htbp]
\centering
\includegraphics[width=0.95\columnwidth]{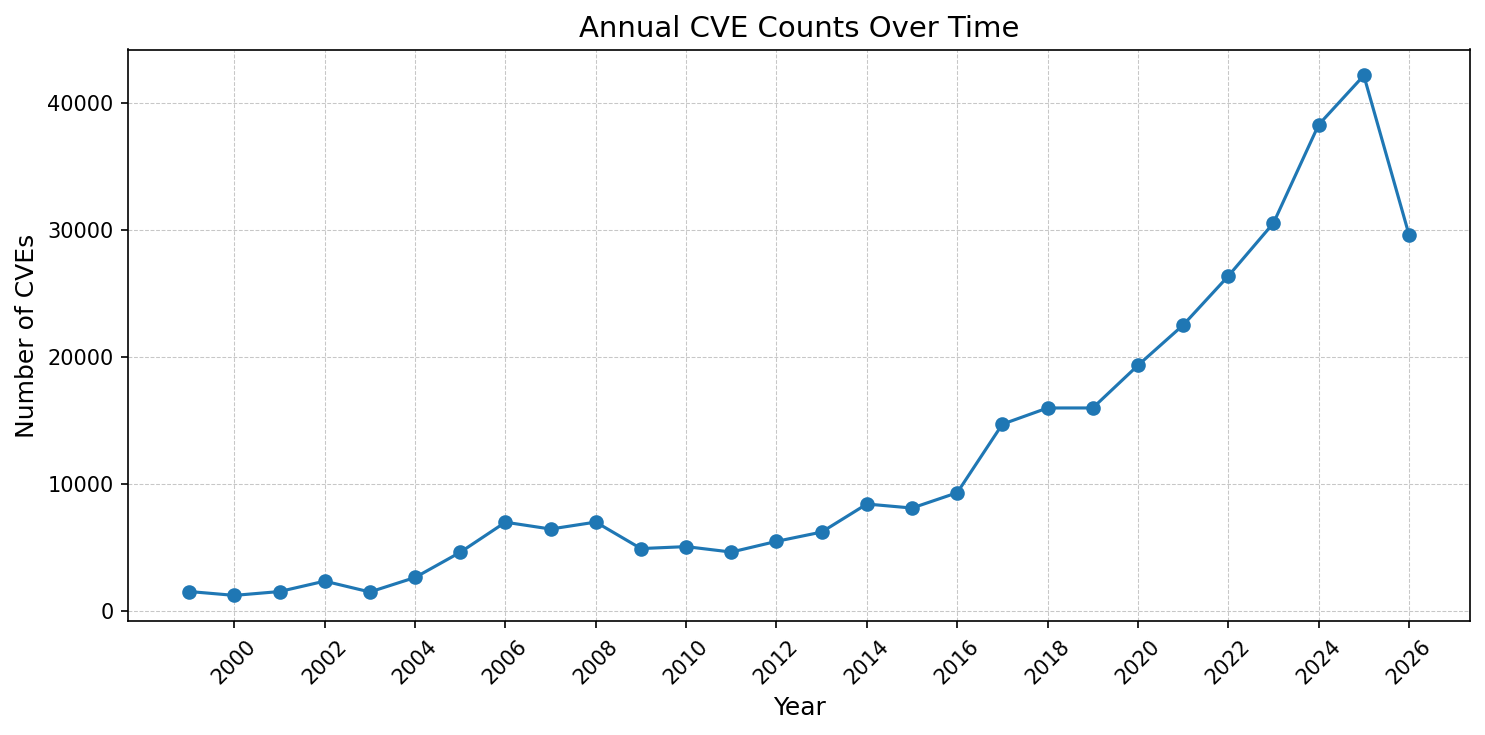}
\caption{Annual number of CVE entries represented in VulLink, grouped by publication year.}
\label{fig:annual_cve_count}
\end{figure}

Fig.~\ref{fig:annual_cve_count} shows the annual distribution of CVE entries represented in VulLink. The graph covers vulnerabilities published from 1999 to 2026 and reflects the rapid growth of vulnerability disclosures in recent years. The lower count for 2026 is expected because the snapshot was collected in early June 2026 and therefore only covers a partial year. This temporal coverage further illustrates the need for a continuously updated graph database, since static snapshots can quickly become incomplete in a fast-changing vulnerability landscape.

\subsection{Embedding Storage and Retrieval Cost}\label{sec:LLMdemission}

VulLink provides pre-computed embeddings of vulnerability descriptions as downloadable semantic features. However, serving high-dimensional embeddings for hundreds of thousands of vulnerability nodes introduces storage, memory, and response-time costs. Therefore, VulLink supports multiple embedding dimensions and uses reduced representations to balance semantic fidelity and system responsiveness.

To guide the dimensionality choices, we measure the storage size and PCA reduction cost of SecBERT embeddings across different dimensions. Table~\ref{tab:secbert-storage-pca} reports the storage cost and PCA transformation cost in terms of time and peak memory usage. The results show a clear trade-off: higher-dimensional embeddings preserve richer semantic information but require more storage and higher processing cost. For example, 32D embeddings require 34MB and less than 25ms for PCA transformation, while 128D embeddings require 133MB and around 125ms. In contrast, 768D embeddings require 795MB and substantially higher transformation cost.

\begin{table}[htbp]
\centering
\caption{SecBERT embedding storage and PCA reduction cost.}
\label{tab:secbert-storage-pca}
\begin{tabular}{|c|c|cc|}
\hline
\textbf{Dimension} & \textbf{Storage (MB)} & \multicolumn{2}{c|}{\textbf{PCA Cost}} \\
\cline{3-4}
 &  & \textbf{Time (ms)} & \textbf{Peak Memory (MB)} \\
\hline
16D   & 17   & 12.9   & 23.89 \\
32D   & 34   & 24.9   & 33.05 \\
64D   & 67   & 53.4   & 66.31 \\
128D  & 133  & 124.5  & 131.92 \\
256D  & 265  & 309.0  & 264.61 \\
512D  & 530  & 921.6  & 268.13 \\
768D  & 795  & 1858.5 & 469.39 \\
\hline
\end{tabular}
\end{table}

Based on these measurements, the deployed system actively serves low- and moderate-dimensional embeddings for interactive use. In particular, 32D embeddings are suitable for lightweight browser-side exploration and quick retrieval, while 128D embeddings provide a practical balance between representation richness and retrieval cost for most API-based downstream tasks. Higher-dimensional embeddings are retained for offline export and evaluation, where users may prefer semantic fidelity over response time. This design supports different user requirements without requiring real-time language-model inference during user queries.

\subsection{Web Interface and API Demonstration}

\begin{figure*}[!t] 
\centering 
\includegraphics[width=1\textwidth]{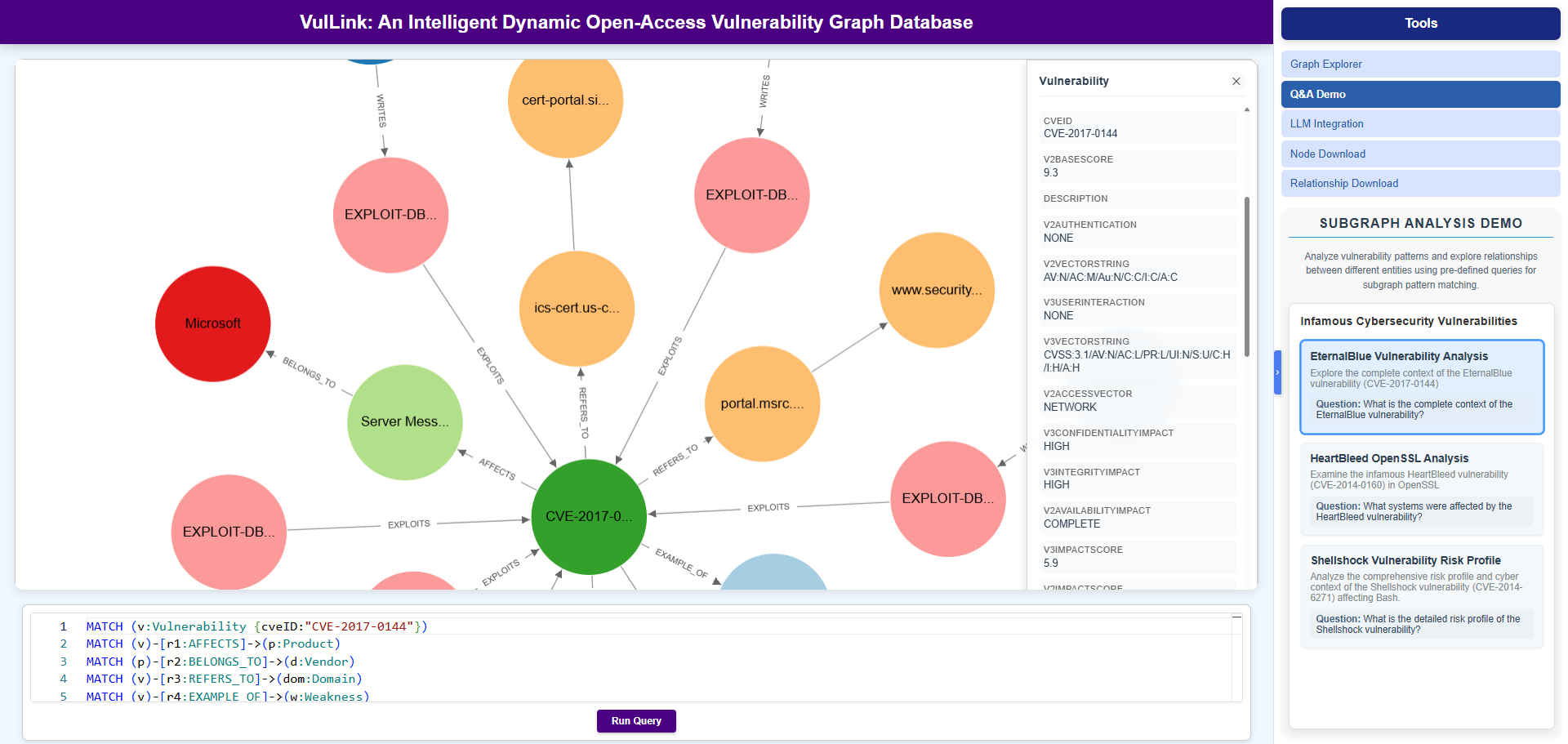} 
\caption{VulLink Web interface with graph visualization, Cypher querying, guided exploration tools, and configurable data retrieval functions.} 
\label{fig:VulLink_visualizer} 
\end{figure*}

Fig.~\ref{fig:VulLink_visualizer} shows the deployed VulLink Web interface. The interface is designed to support both low-barrier exploratory browsing and advanced graph-based data retrieval. The left side contains a graph exploration canvas and an integrated Cypher console, while the right side contains a modular tool panel for guided schema exploration, predefined query demonstrations, embedding retrieval, and configurable data download.

\paragraph{Graph Visualization and Cypher Console}
The graph canvas visualizes the results returned by the Cypher console as an interactive Neo4j-powered node-link diagram. The visualization distinguishes typed cybersecurity entities, such as \emph{Vulnerability}, \emph{Product}, \emph{Weakness}, \emph{Exploit}, \emph{Vendor}, \emph{Author}, and \emph{Domain}, using different colors. Users can zoom in and out, pan across the canvas, drag nodes to adjust the layout, and click nodes or relationships to inspect their properties, such as \texttt{cveID}, CVSS score, description, affected products, exploit metadata, and reference information, in the embedded property panel. The Cypher console can be edited manually by advanced users or automatically populated through the predefined tools in the right-hand panel, enabling filtering, multi-hop traversal, and task-specific subgraph retrieval.

\paragraph{Modular tool panel.}
The right-hand tool panel provides task-oriented functions for both exploratory and technical workflows. Each tool either populates predefined Cypher queries into the console or provides configurable download options for graph and embedding data.
\begin{itemize}
\item \textbf{Graph Explorer:} Provides schema-oriented templates for visualizing node types, relationship types, and common graph patterns. When users click a node or relationship button, the corresponding Cypher query is automatically populated into the console and can be executed to visualize the result in the graph canvas.
\item \textbf{Q\&A Demo:} Provides predefined subgraph analysis examples, such as EternalBlue, HeartBleed, and Shellshock. Selecting an example populates the corresponding Cypher query into the console, allowing users to reproduce the example by clicking \texttt{Run Query}.
\item \textbf{LLM Integration / Embedding Retrieval:} Provides configurable retrieval of pre-computed vulnerability-description embeddings. Users can select the embedding model, publication year, embedding dimension, and file format, and then download the corresponding semantic features for downstream analysis.
\item \textbf{Node Download:} Allows users to select a node type, choose node properties, and download the resulting node table in JSON or CSV format.
\item \textbf{Relationship Download:} Allows users to select a relationship type and download edge data together with source and target node information.
\end{itemize}
This design supports two complementary usage patterns. For exploratory analysis, predefined buttons and examples lower the barrier for users who are unfamiliar with Cypher. For reproducible data mining, configurable download tools allow users to export selected nodes, relationships, subgraphs, and embeddings without rebuilding the data integration pipeline locally.

\paragraph{EternalBlue Multi-hop Query.}
To demonstrate graph-based exploration, we use EternalBlue (CVE-2017-0144), an SMB vulnerability exploited by WannaCry ransomware~\cite{ms17-010} as an example. The following Cypher query retrieves the vulnerability together with its affected products, vendors, reference domains, weakness category, exploits, and exploit authors:
\begin{lstlisting}[basicstyle=\ttfamily\scriptsize] 
MATCH (v:Vulnerability {cveID:"CVE-2017-0144"})
MATCH (v)-[r1:AFFECTS]->(p:Product)
MATCH (p)-[r2:BELONGS_TO]->(d:Vendor)
MATCH (v)-[r3:REFERS_TO]->(dom:Domain)
MATCH (v)-[r4:EXAMPLE_OF]->(w:Weakness)
MATCH (ex:Exploit)-[r5:EXPLOITS]->(v)
MATCH (a:Author)-[r6:WRITES]->(ex)
RETURN v, r1, p, r2, d, r3, dom, r4, w, r5, ex, r6, a
LIMIT 100;
\end{lstlisting}

The resulting subgraph, shown in Fig.~\ref{fig:VulLink_visualizer}, connects the CVE to affected Microsoft products, linked weakness information, ExploitDB records, exploit authors, and external reference domains. This example illustrates how VulLink converts information that is otherwise scattered across repositories into a directly queryable subgraph. The purpose of this demonstration is not to provide a new security analysis of EternalBlue, but to show how the deployed system supports multi-hop graph retrieval for vulnerability investigation and downstream data mining.

\paragraph{Programmatic Data Access} 
VulLink also exposes a public API for automated workflows. The API supports the same core functions as the Web interface: node download, relationship download, customized Cypher queries, and embedding retrieval. Table~\ref{tab:api-endpoints} summarizes the main endpoints.

\begin{table}[htbp]
\caption{Summary of VulLink API endpoints.}
\label{tab:api-endpoints}
\centering
\begin{tabularx}{\columnwidth}{|l|X|}
\hline
\textbf{Endpoint} & \textbf{Description} \\
\hline
docs & Documentation of query parameters and configurations.  \\
\hline
node\_download & Export nodes with selected properties. \\
\hline
relationship\_download & Export edges with source and target node info. \\
\hline
cypher\_query & Submit custom Cypher queries. \\
\hline
llm\_embedding &  Retrieve pre-computed vulnerability-description embeddings by model and dimension. \\
\hline
\end{tabularx}
\end{table}

For example, users can query the \texttt{node\_download} endpoint with parameters such as \texttt{node\_type = "Vulnerability"} and \texttt{props=["cveID", "description", "v3baseScore"]} to obtain structured vulnerability records. Users can also submit Cypher queries through the API to download customized subgraphs, or request embeddings with specified model and dimensionality settings. This programmatic access supports reproducible dataset construction because the same query and parameters can be reused across experiments.

\subsection{Downstream Mining Use Case: Exploitability Prediction}\label{sec:exploitability_prediction}

To evaluate the utility of VulLink for downstream cybersecurity data mining, we conduct an exploitability prediction case study over vulnerability nodes. The goal of this experiment is not to propose a new exploitability prediction model, but to examine whether the graph structure and semantic features provided by VulLink can support downstream machine learning tasks beyond isolated vulnerability attributes.

\paragraph{Task Setup.}

We formulate exploitability prediction as a supervised binary classification task. For each vulnerability node $v \in \mathcal{V}_{\text{vuln}}$, the target label is $y_v \in \{0,1\}$, where $y_v=1$ indicates that the vulnerability is linked to at least one public exploit record in VulLink. In the current graph schema, labels are derived from the existence of an \texttt{EXPLOITS} relationship between an \textit{Exploit} node and the corresponding \textit{Vulnerability} node. This label reflects public exploit evidence available in the integrated ExploitDB records, rather than a claim that the vulnerability is necessarily being actively exploited in the wild.

To incorporate graph context, we construct local heterogeneous subgraphs around vulnerabilities using shared affected products. 
Specifically, we use the shared-product meta-path
\[
\text{Vulnerability}
\xleftarrow{\texttt{AFFECTS}}
\text{Product}
\xrightarrow{\texttt{AFFECTS}}
\text{Vulnerability},
\]
which connects vulnerabilities affected by the same product and allows the graph model to propagate exploitability-related context across product-level neighborhoods, while the non-graph baseline treats each vulnerability as an isolated instance.

\paragraph{Experimental configurations.}
We compare two modeling paradigms. The first is a vulnerability attribute-only baseline using a  Multi-Layer Perceptron (MLP) optimized with batch normalization and dropout ($p=0.3, 0.2$) for standalone binary classification. It operates only on node-level features and does not use graph topology. The second is a graph-augmented representation model using a multi-layer heterogeneous GraphSAGE architecture (HGNN), which performs message passing over the local heterogeneous subgraphs retrieved from VulLink and then applies an MLP classification head. It employs a sum operator ($\bigoplus$) for relation aggregation and a \texttt{NeighborLoader} strategy with $[15,10]$ sampled neighbors per layer.

For both paradigms, we evaluate three feature modalities:
\begin{itemize}
\item \textbf{Embedding Only:} 128-dimensional vulnerability-description embeddings generated from \texttt{SecBERT} and reduced by PCA fitted on the training partition.
\item \textbf{Tabular Only:} structured vulnerability attributes, including reference counts and CVSS v2/v3 base, impact, and exploitability sub-scores, with continuous variables standardized using robust scaling.
\item \textbf{Combined:} concatenation of description embeddings and tabular attributes.
\end{itemize}

\paragraph{Evaluation Protocol.}
We use a stratified random split with 70\% training, 10\% validation, and 20\% testing. Models are optimized using an AdamW optimizer ($lr=1e-3$ for HGNN; $lr=5e-4$ for MLP paired with a Weighted Random Sampler to counter severe label imbalance). We report performance across 10 independent initialization seeds ($S = \{1, 2, \dots, 10\}$) to track statistical significance. Evaluation metrics include Precision-Recall Area Under the Curve (PR-AUC), Macro-averaged F1-Score (Macro-F1), and Precision at top-100 prioritized instances (Precision@100), which simulates practical vulnerability triage constraints faced by cybersecurity operations teams.

\begin{figure}[t]
    \centering
    \includegraphics[width=0.95\columnwidth]{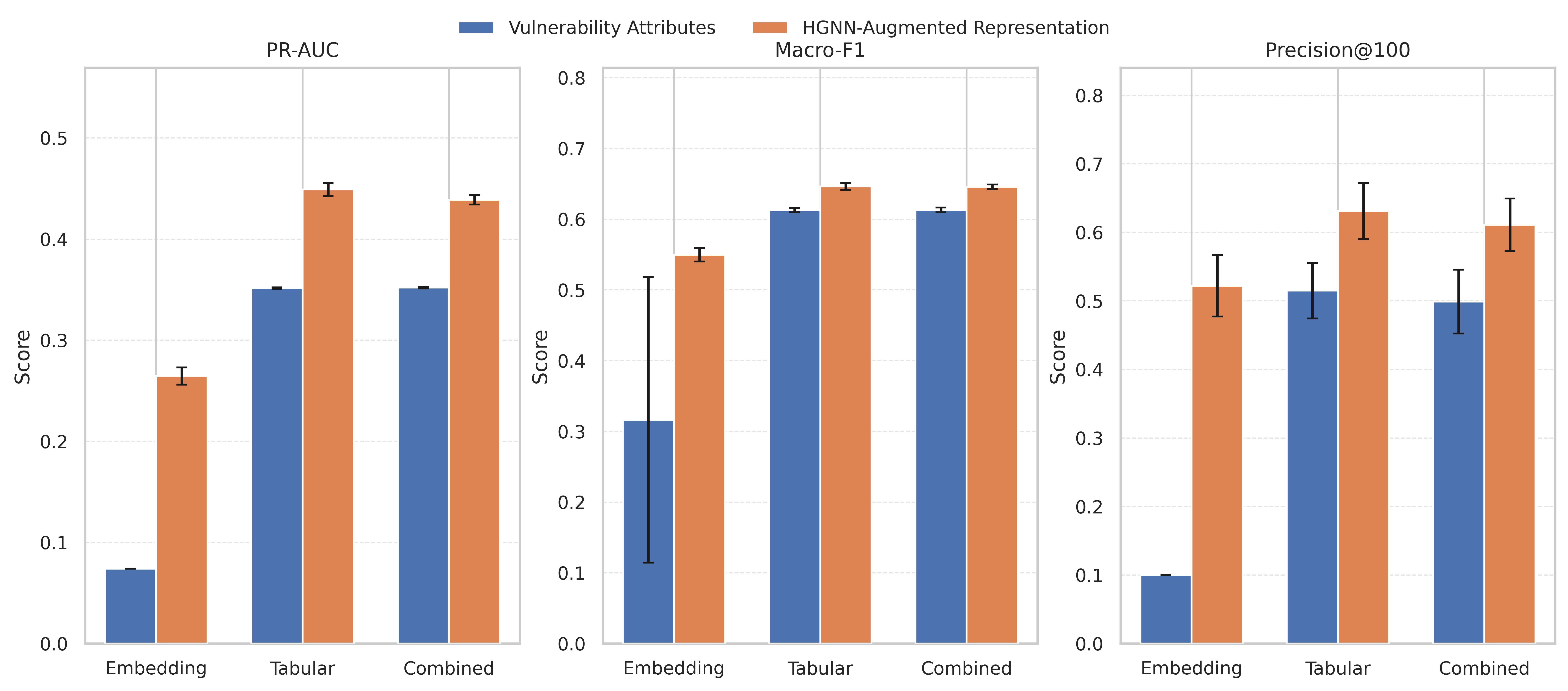}
    \caption{Exploitability prediction results comparing MLP and HGNN across feature modalities.}
    \label{fig:exploitability_comparison}
\end{figure}

\paragraph{Results and Discussion.}
Table~\ref{tab:exploitability_results}  and Fig.~\ref{fig:exploitability_comparison} show that graph-augmented learning consistently improves exploitability prediction across all feature modalities. Compared with the MLP baseline, HGNN improves PR-AUC by 257.8\%, 27.7\%, and 24.7\% under the embedding-only, tabular-only, and combined settings, respectively. Similar improvements are observed for Macro-F1 and Precision@100, indicating that the local graph context retrieved from VulLink provides useful predictive signals beyond isolated node attributes.

\begin{table*}[t]
\centering
\caption{Exploitability prediction performance (\%) across feature modalities. Gain denotes the relative improvement of HGNN over the MLP baseline.}
\label{tab:exploitability_results}
\resizebox{\textwidth}{!}{
\begin{tabular}{l|ccc|ccc|ccc}
\toprule
& \multicolumn{3}{c|}{PR-AUC}
& \multicolumn{3}{c|}{Macro-F1}
& \multicolumn{3}{c}{Precision@100} \\
\cmidrule(lr){2-4}
\cmidrule(lr){5-7}
\cmidrule(lr){8-10}
Feature Modality
& Base (MLP) & HGNN & Gain
& Base (MLP) & HGNN & Gain
& Base (MLP) & HGNN & Gain \\
\midrule
Embedding Only
& 7.39$\pm$0.00
& \textbf{26.45$\pm$0.85}
& \textbf{+257.8\%}
& 31.60$\pm$20.18
& \textbf{54.96$\pm$0.94}
& \textbf{+73.9\%}
& 10.00$\pm$0.00
& \textbf{52.20$\pm$4.49}
& \textbf{+422.0\%}
\\
\midrule
Tabular Only
& 35.15$\pm$0.07
& \textbf{44.89$\pm$0.65}
& \textbf{+27.7\%}
& 61.28$\pm$0.31
& \textbf{64.62$\pm$0.50}
& \textbf{+5.5\%}
& 51.50$\pm$4.06
& \textbf{63.10$\pm$4.11}
& \textbf{+22.5\%}
\\
\midrule
Combined
& 35.19$\pm$0.08
& \textbf{43.87$\pm$0.46}
& \textbf{+24.7\%}
& 61.29$\pm$0.34
& \textbf{64.56$\pm$0.35}
& \textbf{+5.3\%}
& 49.90$\pm$4.66
& \textbf{61.10$\pm$3.83}
& \textbf{+22.4\%}
\\
\bottomrule
\end{tabular}
}
\end{table*}

The largest relative gain appears in the embedding-only setting. Although description embeddings alone perform weakly under the MLP baseline, their performance improves substantially after graph message passing. In particular, HGNN improves the embedding-only setting from 7.39\% to 26.45\% in PR-AUC and from 10.00\% to 52.20\% in Precision@100. Notably, for the triage-oriented Precision@100 metric, the embedding-only HGNN reaches 52.20\%, which is comparable to or slightly higher than the tabular-only MLP baseline of 51.50\%. This suggests that graph propagation can turn otherwise weak textual embeddings into useful signals for top-ranked vulnerability prioritization. However, in terms of overall absolute performance, tabular features remain the strongest modality. The best results are achieved by the HGNN with tabular features, reaching 44.89\% PR-AUC, 64.62\% Macro-F1, and 63.10\% Precision@100. This indicates that structured vulnerability attributes such as CVSS sub-scores and reference counts provide strong predictive signals, and that graph context further enhances their utility.

The combined feature setting does not outperform tabular-only features, suggesting that simple concatenation of textual embeddings and structured attributes is not always beneficial. This result is informative for downstream users of VulLink: the platform provides both semantic and structured graph features, but the optimal feature combination may depend on the task, model, and fusion strategy. Overall, the case study demonstrates that VulLink is not only a graph data repository, but also a practical source of graph context and semantic features for downstream vulnerability mining tasks.

\section{Conclusion}\label{sec:conclusion}
In this paper, we presented VulLink, a deployable, dynamic, and open-access vulnerability graph database for cybersecurity data mining. VulLink integrates multiple public vulnerability repositories through an automated ETL pipeline, converts isolated record-centric data into a continuously updated graph database with typed entities and explicit cross-source relationships, and exposes the resulting resource through an interactive Web interface and public API for graph querying, subgraph export, and embedding download. By providing pre-computed vulnerability-description embeddings generated by pretrained language models, VulLink further supports downstream mining tasks that combine structured vulnerability attributes, graph context, and semantic features. The deployed system currently contains over 340,000 vulnerability nodes and more than 1.6 million relationships, and the exploitability prediction case study demonstrates its practical utility for downstream machine learning analysis. Future work will focus on improving scalability under continuous graph growth, integrating additional vulnerability and threat intelligence sources, and extending graph mining utilities for tasks such as dependency-aware risk analysis and graph-evidence-grounded patch prioritization.

\bibliographystyle{IEEEtran}
\bibliography{references}
\end{document}